\documentclass[conference]{IEEEtran}
\IEEEoverridecommandlockouts
\usepackage{cite}
\usepackage{amsmath,amssymb,amsfonts}
\usepackage{algorithmic}
\usepackage{graphicx}
\usepackage{textcomp}
\usepackage{xcolor}
\usepackage{algorithm}

\def\BibTeX{{\rm B\kern-.05em{\sc i\kern-.025em b}\kern-.08em
		T\kern-.1667em\lower.7ex\hbox{E}\kern-.125emX}}
\begin{document}
	
	\title{Fluid-Antenna Enhanced Integrated \\
		Sensing and Communication: Joint Antenna Positioning and Beamforming Design}
	
	\author{\IEEEauthorblockN{Tian Hao$^*$, Changxin Shi$^*$, Yinghong Guo$^*$, Bin Xia$^*$, Feng Yang$^*$}
		\IEEEauthorblockA{\textit{$^*$Institute of Wireless Communication Technologies, Shanghai Jiao Tong University, Shanghai, China}\\
			{Email:\{optimization, shichx, yinghongguo, bxia, yangfeng\}}@sjtu.edu.cn}
	}
	
	\maketitle
	
	\begin{abstract}
	
	
	This paper investigates a fluid antenna (FA) enhanced integrated sensing and communication (ISAC) system consisting of a base station (BS), multiple single-antenna communication users, and one point target, where the BS is equipped with FAs to enhance both the communication and sensing performance. First, we formulate a problem that maximizes the radar signal-to-noise ratio (SNR) by jointly optimizing the FAs' positions and transmit beamforming matrix. Then, to tackle this highly non-convex problem, we present efficient algorithms by using alternating optimization (AO), successive convex approximation (SCA), and semi-definite relaxation (SDR). Numerical results demonstrate the convergence behavior and effectiveness of the proposed algorithm. 
	\end{abstract}

	\let\thefootnote\relax\footnotetext{This work was supported in part by the Natural Science Foundation of Hunan Province under Grant 2024JJ8023, and in part by the Shanghai Municipal Science and Technology Major Project under Grant 2021SHZDZX0102.}
	\section{Introduction}
	With the explosive growth in the number of communication and sensing devices, spectrum resources are under increasing pressure. As a promising solution, integrated sensing and communication (ISAC) is recently proposed to tackle this challenge by deploying both communication and sensing on a hardware platform while sharing the spectrum of the radar system and communication system \cite{b1}. 
	Compared to the communication-only and sensing-only systems, which operate independently in different frequency bands and platforms, ISAC can reduce hardware costs and reuse spectrum resources\cite{b2,b3,b4}. Therefore, ISAC has attracted much interest from industry and academia.

	Multiple-input multiple-output (MIMO) technology is widely used in ISAC systems to increase the number of spatial degrees of freedom, thereby improving radar perception performance and realizing multi-user communication \cite{b5}. However, due to the fixed position of antennas in the traditional MIMO-ISAC system, the diversity and spatial multiplexing performance are limited. Recently, fluid antenna (FA) is developed as a promising solution to further exploit the wireless channel variation in the continuous spatial domain \cite{b6,b7,b8}. Specifically, in FA-enabled systems, each antenna element is connected to an RF chain via a flexible cable, and the position of FA can be adjusted in real time. Thus, the channel can be configured to improve the system performance by adjusting the locations of multiple fluid antennas. 

	Existing works propose various FA-based optimization schemes to verify the superior performance of the FA-assisted system compared with the traditional fixed-position antenna (FPA)-based system \cite{b9,b10,b11,b12}. In \cite{b9}, the authors consider a point-to-point MIMO communication system where the transmitter and receiver are both equipped with FAs and maximize the channel capacity by jointly optimizing the antenna positions of the receiver and the transmitter. In \cite{b10}, the authors consider the uplink transmission where BS and users are equipped with FPAs and single-FA, and minimize the total transmit power at the BS by jointly  optimizing the antenna positions, the transmit power at each user, and the receive beamforming matrix. In \cite{b11}, the authors jointly optimize the antenna positions, transmit and receive beamforming, and power allocation to maximize the rate. In \cite{b12}, the authors combine reconfigurable intelligent surface (RIS) and FA systems to protect against multi-user interference at each user device. So far, most of the works on FA focus on how to design FAs' positions to improve the performance of the communication system. To the best of our knowledge, the FA-enhanced ISAC system has not been studied thus far.

	Inspired by the above, this paper studies the FA-enhanced integrated sensing and communication system. In particular, the base station (BS) is equipped with FAs as transmitting antennas and with FPAs as receiving antennas, which serve multiple users while simultaneously sensing one target. Under the constraints of the finite moving region of FAs, the minimum FA distance, and the minimum signal-to-interference-plus-noise (SINR) per user,  we propose an optimization problem to maximize the radar signal-to-noise (SNR) by jointly optimizing the positions of FAs and transmit beamforming matrix at the BS. Since the resulting problem is non-convex and involves highly coupling variables, we propose an iterative alternating optimization (AO) \cite{b9} algorithm based on successive convex approximation (SCA) \cite{b14} to obtain a sub-optimal solution. The simulation results show that the proposed FA-based ISAC can significantly improve the sensing ability while ensuring the communication quality of service (QoS) through antenna position optimization.
	
	\section{System Model and Problem Formulation}
	\subsection{FA-Enabled ISAC System}
	We consider an ISAC system where the BS communicates with $K$ communication users (CU), each of which is equipped with a single fixed position antenna (FPA), while sensing a point target by a dual-functional signal. For convenience, let $\mathcal{K}\triangleq \left\{1,2,...,K \right\}$ denote the set of users indices. The BS is equipped with a planer array with $N_t$ fluid antennas (FAs) for signal transmission and a uniform planer array (UPA) consisting of $P \times Q$ FPAs for signal reception. The channel condition can be configured by changing the FAs' position in a rectangular region $\mathcal{C}_t$ of size $W\times L$ ($m \times m$) \cite{b9}, the center of which is denoted as the origin point of $\mathcal{C}_t$. The position of the $m$-th FA is represented by its coordinates on $\mathcal{C}_t$, i.e.,  $\mathbf{t}_m\triangleq[x_m,y_m]^T\in\mathcal{C}_t$.
	
	\begin{figure}[htbp]
		\centering{\includegraphics[width=1\linewidth]{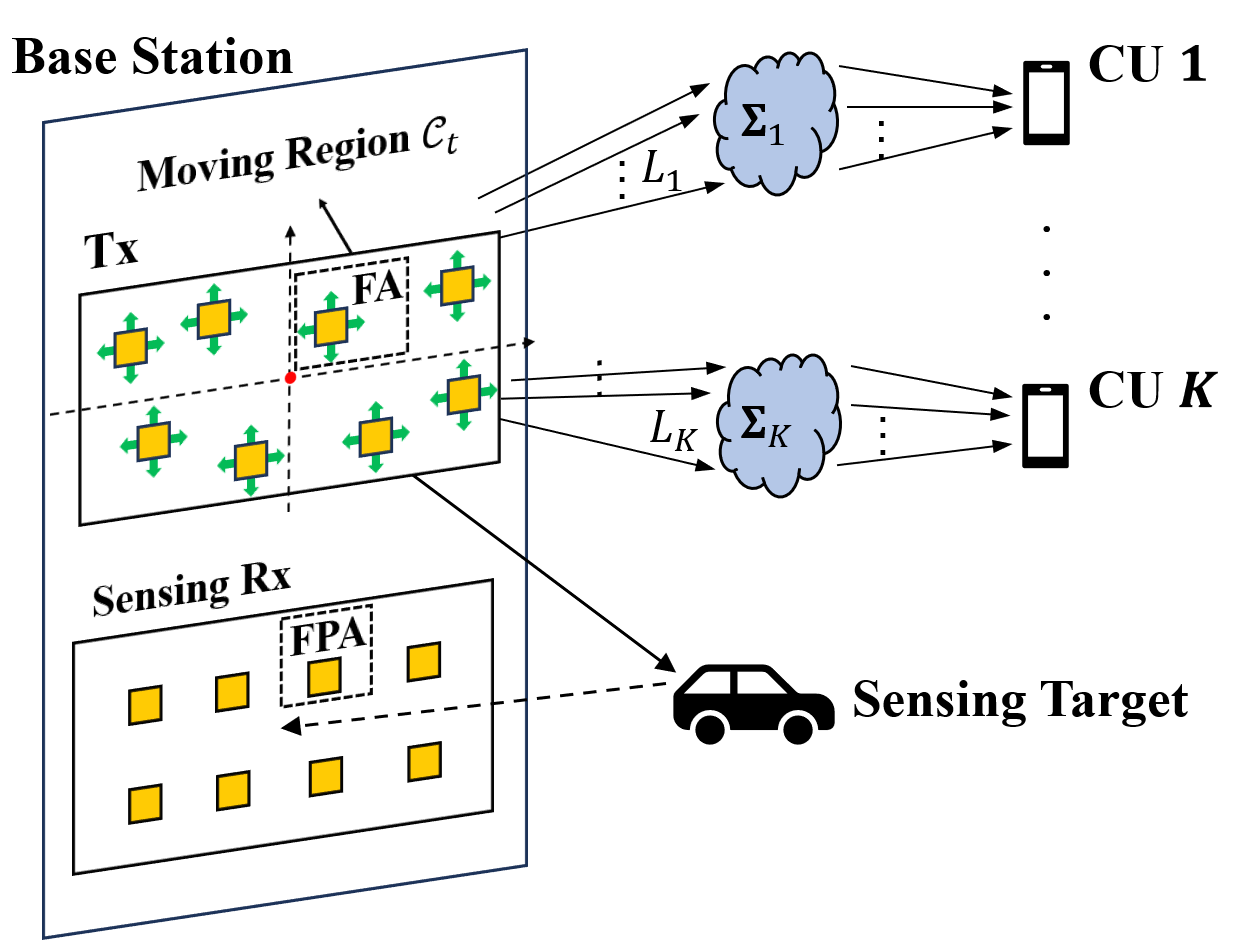}}
		\caption{ The FA-aided ISAC system.}
		\label{systemmodel}
	\end{figure} 
	Let $\mathbf{w}_k\in \mathbb{C}^{N_t\times 1}$ represent the beamforming for user $k$. The signal $\mathbf{x}\in \mathbb{C}^{N_t\times 1}$ transmitted by ISAC BS can be written as:
	\begin{equation} 
		\mathbf{x}\triangleq\sum_{k=1}^{K}\mathbf{w}_k s_k=\mathbf{W}\mathbf{s},
	\end{equation}
	where $\mathbf{s}=[s_1,...,s_K]^T\in \mathbb{C}^{K\times 1}$  with $\mathbb{E}[\mathbf{s}\mathbf{s}^H]=\mathbf{I}_K$ represents the data symbols for the communication users, and $\mathbf{W}=[\mathbf{w}_1,...,\mathbf{w}_K]\in \mathbb{C}^{N_t\times K}$. Note that $\mathbf{x}$ is the dual-functional signal, which can be used for both sensing and communication \cite{b5}. Then, we have the following constraint for the transmit power of the BS:
	\begin{equation}
		E(\Vert \mathbf{x} \Vert_2^2) \triangleq \text{Tr}(\mathbf{W}\mathbf{W}^H)\leq P_{max},\label{power_cons}
	\end{equation}
	where $P_{max}$ represents the maximum available power budget at the BS.
	
	\subsection{Communication Model}
	We adopt a far field-response based channel model for communication channel. Since the sizes of moving region for the FAs are much smaller than the signal propagation distance of each path between BS and CU, the angle-of-arrival (AoA), the angle-of-departure (AoD) and the amplitude of path response remain constant at different positions in FAs. Let $L_k$ denote the number of paths between the BS and the user $k$. For user $k$, the signal propagation difference for the $l$-th channel path between the position of $m$-th FA and the origin point the moving region is written as 
	\begin{align}
		&\rho(\mathbf{t}_m,\theta_{k,l},\phi_{k,l})=x_m cos\theta_{k,l}sin\phi_{k,l} + y_msin\theta_{k,l}, \nonumber \\
		&\qquad \qquad \qquad \qquad \qquad \qquad \qquad k\in \mathcal{K}, 1\leq l \leq L_k,
	\end{align}
	where $\theta_{k,l}$ and $\phi_{k,l}$ denote the elevation and azimuth AoDs of the $l$-th path of channel between the user $k$ and the BS. Accordingly, the transmit field response vector between $k$-th user and $m$-th FA at BS is given by:
	\begin{equation}
		\mathbf{g}_k(\mathbf{t}_m)=[e^{j\frac{2\pi}{\lambda}\rho(\mathbf{t}_m,\theta_{k,1},\phi_{k,1})},...,e^{j\frac{2\pi}{\lambda}\rho(\mathbf{t}_m,\theta_{k,L_k},\phi_{k,L_k})}]^T,
	\end{equation} 
	where $\lambda$ is the carrier wavelength. Thus, the communication channel between the BS and user $k$ is modeled as follows \cite{b10}:
	\begin{equation}
		\mathbf{h}_k(\tilde{\mathbf{t}})= \mathbf{1}^T_{L_k}\mathbf{\Sigma}_k \mathbf{G}_k(\tilde{\mathbf{t}})\in \mathbb{C}^{N_t\times 1}, \label{channel}
	\end{equation}
	where $\tilde{\mathbf{t}}=[\mathbf{t}_1^T,...,\mathbf{t}_{N_t}^T]^T\in \mathbb{R}^{2N_t \times 1}$ represents the FAs position, $\mathbf{G}_k(\tilde{\mathbf{t}})=[\mathbf{g}_k(\mathbf{t}_1),\mathbf{g}_k(\mathbf{t}_2),...,\mathbf{g}_k(\mathbf{t}_{N_t})]\in \mathbb{C}^{L_k\times N_t}$ represents the field response matrix at the BS, and $\mathbf{\Sigma}_k=diag \{ [\sigma_{1,k},...,\sigma_{L_k,k}]^T \}$ denotes path response of $L_k$ paths of the channel between the BS and the $k$-th user. 
	
	For the downlink communication, we adopt the per-user signal-to-interference-plus-noise ratio (SINR) to measure the communication quality-of-service. The received signal of $k$-th user is given by:
	\begin{equation}
		y_k = \mathbf{h}_k(\tilde{\mathbf{t}}) \mathbf{x}+z_k, k\in\mathcal{K},
	\end{equation}
	where $\mathbf{h}_k(\tilde{\mathbf{t}})$ is given by (\ref{channel}), $z_k \sim \mathcal{CN}(0,\sigma_r^2)$ denotes the zero-mean additive white Gaussian noise (AWGN) with variance $\sigma^2$. Then, the SINR of user $k$ is given by:
	\begin{equation}
		\gamma_k(\mathbf{W},\tilde{\mathbf{t}}) = \frac{|\mathbf{h}_k(\tilde{\mathbf{t}}) \mathbf{w}_k|^2}{\sum_{q=1,q\neq k}^{K}|\mathbf{h}_k(\tilde{\mathbf{t}}) \mathbf{w}_q|^2 + \sigma^2},
	\end{equation}
	
	\subsection{Sensing Model}
	Let $\vartheta$ and $\varphi$ denote the elevation and azimuth angle between the target and the BS, respectively. We adopt the line-of-sight (LoS) channel model for the sensing channel between the BS and the target \cite{b5}. Since the relative far distance between the BS and the point target, $\vartheta$ and $\varphi$ are constant at different positions in FAs. Let $\mathbf{a}_r(\vartheta,\varphi)\in \mathbb{C}^{1\times PQ}$ and $\mathbf{a}_t(\vartheta,\varphi,\tilde{\mathbf{t}})\in \mathbb{C}^{1\times N_t}$ denote the receive and transmit steering vectors. Specially, the receive steering vector $\mathbf{a}_r(\vartheta,\varphi)$ is denoted by 
	\begin{equation}
		\mathbf{a}_r(\vartheta,\varphi) = \mathbf{a}_P(\vartheta,\varphi) \otimes \mathbf{a}_Q(\vartheta,\varphi),
	\end{equation}
	where $\mathbf{a}_P(\vartheta,\varphi) = \left[1, e^{j\pi cos\vartheta sin\varphi},...,e^{j\pi(P-1)cos\vartheta sin\varphi} \right]$, $ \mathbf{a}_Q(\vartheta,\varphi) = \left[1, e^{j\pi sin\vartheta},...,e^{j\pi(Q-1)sin\vartheta} \right]$. The transmit steering vector, i.e.,  $\mathbf{a}_t(\theta,\phi,\tilde{\mathbf{t}})$ can be written as
	\begin{equation}
		\mathbf{a}_t(\vartheta,\varphi,\tilde{\mathbf{t}}) = \left[ e^{j\frac{2\pi}{\lambda}\rho(\mathbf{t}_1,\vartheta,\varphi)},...,e^{j\frac{2\pi}{\lambda}\rho(\mathbf{t}_{N_t},\vartheta,\varphi)} \right].
	\end{equation}
	Then, the sensing channel $\mathbf{G}$ can be written as \cite{b5}
	\begin{equation}
		\mathbf{G} \triangleq \alpha \mathbf{a}_r(\vartheta,\varphi)^H 	\mathbf{a}_t(\vartheta,\varphi,\tilde{\mathbf{t}}),  
	\end{equation}
	where $\alpha$ denotes the reflection coefficient of the sensing target. Thus, for radar sensing, the reflected echo signal $\mathbf{y}_r \in \mathbb{C}^{PQ \times 1}$ at the sensing receiver of the BS can be written as :
	\begin{equation}
		\mathbf{y}_r=\mathbf{G}\mathbf{x}+\mathbf{z}_r,
	\end{equation}
	where $\mathbf{z}_r\in \mathbb{C}^{PQ \times 1}$ is the AWGN following $\mathcal{CN}(0,\sigma_r^2\mathbf{I}_{PQ})$. Subsequently, the sensing SNR at the sensing receiver which can be written as
	\begin{align}
		& \gamma_s(\mathbf{W},\tilde{\mathbf{t}},\vartheta,\varphi) = \eta \mathbf{a}_t(\vartheta,\varphi,\tilde{\mathbf{t}}) \mathbf{W} \mathbf{W}^H \mathbf{a}_t(\vartheta,\varphi,\tilde{\mathbf{t}})^H, \label{sensing_SNR}
	\end{align}
	where $\eta\triangleq\frac{|\alpha|^2 \mathbf{a}_{r}(\vartheta,\varphi)\mathbf{a}_{r}^H(\vartheta,\varphi)}{\sigma_r^2}$ denotes the integrated coefficient.
	
	\section{Problem Formulation}
	In this paper, we aim to maximize the radar SNR while satisfying the SINR requirements of users by jointly optimizing the FA position $\tilde{\mathbf{t}}$ and the transmit beamforming matrix $\mathbf{W}$. Depending on whether perfect CSI of the communication channels and the target location are available at the BS, we consider two scenarios elaborated as below
	
	Then the corresponding optimization problem is formulated as
	\begin{align}
		(\text{P1})~~\underset{\mathbf{W},\tilde{\mathbf{t}}}{\text{max}}&~\gamma_r(\mathbf{W},\tilde{\mathbf{t}}) \label{Problem} \\
		\text{s.t.}&~\frac{|\mathbf{h}_k(\tilde{\mathbf{t}}) \mathbf{w}_k|^2}{\sum_{q=1,q\neq k}^{K}|\mathbf{h}_k(\tilde{\mathbf{t}}) \mathbf{w}_q|^2 + \sigma^2} \geq \Gamma_k,~\forall k \tag{\ref{Problem}{a}} \label{Problema}\\
		&~\mathbf{t}_m\in \mathcal{C}_t,~1\leq m \leq N_t, \tag{\ref{Problem}{b}} \label{Problemb}\\
		&~\left\| \mathbf{t}_m - \mathbf{t}_n \right\|_2^2 \geq D^2,~1\leq m \neq n \leq N_t, \tag{\ref{Problem}{c}} \label{Problemc}\\
		&~\sum_{k=1}^{K}\text{Tr}(\mathbf{w}_k\mathbf{w}_k^H)\leq P_{max}, \tag{\ref{Problem}{d}} \label{Problemd}
	\end{align}
	where the constraint (\ref{Problema}) denotes the minimum SINR requirement of each user, the constraint (\ref{Problemb}) indicates the movement region of each FA, and the constraint (\ref{Problemc}) guarantees the minimum distance $D$ between each pair of FAs.
	
	Note that problem (\ref{Problem}) is non-convex because the objective function is non-concave over the FA position $\tilde{\mathbf{t}}$, and the constraints (\ref{Problema}) and (\ref{Problemc}) are non-convex. Besides, the FA position $\tilde{\mathbf{t}}$ and transmit beamforming matrix $\mathbf{W}$ are highly coupled in objective function and constraint (\ref{Problema}), which makes (P1) challenging to solve.
	\section{Proposed Algorithm}
	In this section, we first propose an efficient algorithm to solve (P1) by applying AO and SCA. Specifically, we decompose (P1) into two sub-problems. For given FA position $\tilde{\mathbf{t}}$, we optimize the beamforming matrix $\mathbf{W}$ at the transmitter. For given beamforming matrix $\mathbf{W}$, the FA position $\tilde{\mathbf{t}}$ is optimized based on successive convex optimization techniques. Finally, we present the overall algorithm than analyze its convergence. 
	
	\subsection{Optimization of Transmit Beamforming}
	For any given FA position $\tilde{\mathbf{t}}$, the transmit beamforming can be obtained by solving the following problem
	\begin{align}
		(\text{P2})~~\underset{\mathbf{W}}{\text{max}}&~\eta \mathbf{a}_t(\vartheta,\varphi,\tilde{\mathbf{t}}) \mathbf{W} \mathbf{W}^H \mathbf{a}_t(\vartheta,\varphi,\tilde{\mathbf{t}})^H \label{Problem2} \\
		\text{s.t.}&~(\ref{Problema}), (\ref{Problemd}) \nonumber.
	\end{align}
	Notice that problem ($\ref{Problem2}$) is non-convex due to the non-convex constraints ($\ref{Problema}$). We then employ the semidefinite relaxation (SDR) technique to solve the problem ($\ref{Problem2}$). 
	
	Let $\mathbf{W}_k\triangleq\mathbf{w}_k\mathbf{w}_k^H$, $\mathbf{H}_k\triangleq\mathbf{h}_k^H\mathbf{h}_k$, $\mathbf{A}\triangleq\eta\mathbf{a}_t(\vartheta,\varphi,\tilde{\mathbf{t}})^H\mathbf{a}_t(\vartheta,\varphi,\tilde{\mathbf{t}})$. Thus, (P2) can be approximated as
	\begin{align}
		(\text{P3})~~\underset{\mathbf{W}_k}{\text{max}}&~ \text{Tr}(\mathbf{A}\mathbf{W}_k) \label{Problem3} \\
		\text{s.t.}&\text{Tr}(\mathbf{H}_k\mathbf{W}_k)-\Gamma_k\sum_{q=1,\neq k}^{K}\text{Tr}(\mathbf{H}_k\mathbf{W}_q) \geq \Gamma_k\sigma^2, \forall k,  \nonumber\\
		&~\sum_{k=1}^{K}\text{Tr}(\mathbf{W}_k) \leq P_{max}, \nonumber \\
		&~\mathbf{W}_k \succeq \mathbf{0}, \forall k , \nonumber \\
		&~\text{rank}(\mathbf{W}_k)=1,~\forall k  \tag{\ref{Problem3}{a}} \label{Problem3a}\nonumber.
	\end{align}
	
	Then, by dropping the rank-1 constraints ($\ref{Problem3a}$), problem (\ref{Problem3}) becomes a standard semidefinite program (SDP), which can be solved via CVX \cite{b13}. It is noteworthy that the solution obtained from a non-rank-1 SDR problem can be guaranteed to be rank-1 \cite{b5}. With $\mathbf{W}_k^*$ denoting the solution of the non-rank-1 problem, the optimal beamformer can be obtained by the following formula:
	\begin{equation}
		\mathbf{w}_k^* =\sqrt{\lambda_{max}(\mathbf{W}_k^*)}\mathcal{P}(\mathbf{W}_k^*), \label{decomp}
	\end{equation}
	where $\lambda_{max}(\mathbf{W}_k^*)$ represents the largest eigenvalue of $\mathbf{W}_k^*$, and $\mathcal{P}(\mathbf{W}_k^*)$ is the eigenvector of $\lambda_{max}(\mathbf{W}_k^*)$.
	
	\subsection{FA Position Optimization}
	For any given transmit beamforming matrix $\mathbf{W}$, the FA position can be obtained by solving the following problem.
	\begin{align}
		(\text{P4})~~\underset{\tilde{\mathbf{t}}}{\text{max}}&~ \mathbf{a}_t(\vartheta,\varphi,\tilde{\mathbf{t}}) \mathbf{Q} \mathbf{a}_t(\vartheta,\varphi,\tilde{\mathbf{t}})^H \label{Problem4} \\
		\text{s.t.}&~\frac{|\mathbf{h}_k(\tilde{\mathbf{t}}) \mathbf{w}_k|^2}{\sum_{q=1,q\neq k}^{K}|\mathbf{h}_k(\tilde{\mathbf{t}}) \mathbf{w}_q|^2 + \sigma^2} \geq \Gamma_k,~\forall k \tag{\ref{Problem4}{a}} \label{Problem4a}\\
		&~\mathbf{t}_m\in \mathcal{C}_t,~1\leq m \leq N_t, \tag{\ref{Problem4}{b}} \label{Problem4b}\\
		&~\left\| \mathbf{t}_m - \mathbf{t}_n \right\|_2^2 \geq D^2,~1\leq m \neq n \leq N_t, \tag{\ref{Problem4}{c}} \label{Problem4c} 
	\end{align}
	where $\mathbf{Q}\triangleq\eta\mathbf{W}\mathbf{W}^H=\{q_{i,j}=|q_{i,j}|e^{j\varphi_{i,j}},1\leq i,j\leq N_t\}$. Problem (\ref{Problem4}) is non-convex due to the non-concave objective function and the non-convex constraints in (\ref{Problem4a}) and (\ref{Problem4c}). It is challenging to solve problem (\ref{Problem4}). In the following, we tackle (P4) using the SCA for the FA position optimization. 
	
	To solve the non-convexity of objective function, (\ref{Problem4a}) and (\ref{Problem4c}), successive convexity optimization technique can be applied, where in each iteration, the original function is approximated by a more tractable function at a given local point. According to Taylor's theorem, we construct a surrogate function by using second-order Taylor expansion that locally approximates the objective function $g(\tilde{\mathbf{t}})$ \cite{b14}. Specifically, let $\tilde{\mathbf{t}}^r$ represent the given position of FAs in the $r$-th iteration. The objective function is globally lower-bound by a concave function over $\tilde{\mathbf{t}}$, which is denoted by:
	\begin{equation}
		g(\tilde{\mathbf{t}})\geq g(\tilde{\mathbf{t}}^r)+\nabla g(\tilde{\mathbf{t}})^T(\tilde{\mathbf{t}}-\tilde{\mathbf{t}}^r)-\frac{\delta}{2}(\tilde{\mathbf{t}}-\tilde{\mathbf{t}}^r)^T(\tilde{\mathbf{t}}-\tilde{\mathbf{t}}^r). \label{obj_ste}
	\end{equation}
	where $g(\tilde{\mathbf{t}})\triangleq \mathbf{a}_t(\vartheta,\varphi,\tilde{\mathbf{t}}) \mathbf{Q} \mathbf{a}_t(\vartheta,\varphi,\tilde{\mathbf{t}})^H$, $\nabla g(\tilde{\mathbf{t}}) $ and $\nabla^2 g(\tilde{\mathbf{t}})$ represent the gradient vector and the Hessian matrix of $g(\tilde{\mathbf{t}})$ on $\tilde{\mathbf{t}}$. The derivatives of $\nabla g(\tilde{\mathbf{t}}) $ and $\nabla^2 g(\tilde{\mathbf{t}})$ are given in Appendix. Note that $\Vert \nabla^2 g(\tilde{\mathbf{t}}) \Vert_2^2 \leq \Vert \nabla^2 g(\tilde{\mathbf{t}}) \Vert_F^2$ and $\Vert \nabla^2 g(\tilde{\mathbf{t}}) \Vert_2\mathbf{I}_{2N_t} \succeq \nabla^2 g(\tilde{\mathbf{t}})$, we construct a positive real number $\delta$ based on $\Vert \nabla^2 g(\tilde{\mathbf{t}}) \Vert_F$ such that $\delta \mathbf{I}_{2N_t}\succeq \nabla^2 g(\tilde{\mathbf{t}})$. Depending on the derivatives of $\nabla^2 g(\tilde{\mathbf{t}})$ in Appendix, the $\delta$ is given by:
	
	\begin{equation}
		\delta = \frac{16N_t \pi^2 N_t\sqrt{N_t-1}\epsilon}{\lambda^2},
	\end{equation}
	where $\epsilon = max(|q_{i,j}|)$.
	
	For constraints (\ref{Problem4c}), since $\left\| \mathbf{t}_m - \mathbf{t}_n \right\|_2^2$ is a convex function with respect to $\mathbf{t}_m - \mathbf{t}_n$, we get the following inequality by applying the first-order Taylor expansion at the given point $\mathbf{t}_m^r - \mathbf{t}_n^r$
	\begin{align}
		\left\| \mathbf{t}_m - \mathbf{t}_n \right\|_2^2 \geq -\left\| \mathbf{t}_m^r - \mathbf{t}_n^r \right\|_2^2 + 2(\mathbf{t}_m^r - \mathbf{t}_n^r)^T  \nonumber \\
		\times (\mathbf{t}_m - \mathbf{t}_n),~1\leq m \neq n \leq N_t. \label{C3_bound}
	\end{align}
	
	For constraint (\ref{Problem4a}),  it can be transformed into the following form:
	\begin{equation}
		\mathbf{h}_k(\tilde{\mathbf{t}}) \mathbf{R}_k \mathbf{h}_k^H(\tilde{\mathbf{t}}) + \Gamma_k \sigma^2 \leq 0, \forall k,
	\end{equation}
	where $\mathbf{R}_k=\sum_{q=1,q\neq k}^{K}\mathbf{w}_q\mathbf{w}_q^H-\mathbf{w}_k\mathbf{w}_k^H$. The term $\mathbf{h}_k(\tilde{\mathbf{t}}) \mathbf{R}_k \mathbf{h}_k^H(\tilde{\mathbf{t}})$ can be transformed to ($\ref{long_eq}$), where we define:
	\begin{equation}
		\kappa_{k,i,j,l,p}\triangleq\frac{2\pi}{\lambda}(\rho_{k,l}(\mathbf{t}_i)-\rho_{k,p}(\mathbf{t}_j))+\angle \sigma_{l,k} -\angle \sigma_{p,k}+ \angle \mathbf{R}_k(i,j), 
	\end{equation}
	\begin{equation}
		\mu_{k,i,j,l,p}\triangleq|\sigma_{l,k}||\sigma_{p,k}||\mathbf{R}_k(i,j)|,
	\end{equation}
	\begin{figure*}[ht]
		\normalsize
		\begin{equation}
			f_k(\tilde{\mathbf{t}}) = \sum_{i=1}^{N_t}\sum_{l=1}^{L_k}|\sigma_{l,k}|^2\mathbf{R}_k(i,i)
			+\sum_{i=1}^{N_t}\sum_{l=1}^{L_k-1}\sum_{p\neq l}^{L_k}2\mu_{k,i,i,l,p}cos(\kappa_{k,i,i,l,p})
			+\sum_{i=1}^{N_t-1}\sum_{j=i+1}^{N_t}\sum_{l=1}^{N_t}\sum_{p=1}^{N_t}\mu_{k,i,j,l,p}cos(\kappa_{k,i,j,l,p}) \label{long_eq}
		\end{equation}	
		\hrulefill
		\vspace*{4pt}
	\end{figure*}
	As $f_k(\tilde{\mathbf{t}})$ is neither convex nor concave with respect to $\tilde{\mathbf{t}}$, we construct a surrogate function that serves as an upper bound of $f_k(\tilde{\mathbf{t}})$ by using the second-order Taylor expansion. We apply second-order Taylor expansion to $f_k(\tilde{\mathbf{t}})$ around $\tilde{\mathbf{t}}^r$:
	\begin{equation}
		f_k(\tilde{\mathbf{t}}) \leq f_k(\tilde{\mathbf{t}}^r) + \nabla f_k(\tilde{\mathbf{t}}^r)^T (\tilde{\mathbf{t}} - \tilde{\mathbf{t}}^r) +\frac{\zeta_k}{2}(\tilde{\mathbf{t}} - \tilde{\mathbf{t}}^r)^T(\tilde{\mathbf{t}} - \tilde{\mathbf{t}}^r), \label{C4_bound}
	\end{equation}  
	where $ \nabla f_k(\tilde{\mathbf{t}}^r)$ denotes the gradient vector of $f_k(\tilde{\mathbf{t}})$, $\zeta_k$ is a positive real number which satisfies $\zeta_k \mathbf{I}_{2N_t}\succeq \nabla^2 f_k(\tilde{\mathbf{t}})$. $\nabla f_k(\tilde{\mathbf{t}}^r)$ and $\nabla^2 f_k(\tilde{\mathbf{t}})$ can be obtained according to the calculation method of $\nabla g(\tilde{\mathbf{t}}^r)$ and $\nabla^2 g(\tilde{\mathbf{t}})$ in the Appendix. Similar to the calculation of $\delta$, $\zeta_k$ can be given by:
	\begin{align}
		&~\zeta_k = \frac{16N_t \pi^2}{\lambda^2} ( \sum_{l=1}^{L_k-1}\sum_{p=l}^{L_k}|\sigma_{l,k}||\sigma_{p,k}|\eta_k  \nonumber \\
		&~+ \sum_{l=1}^{L_k}\sum_{p=1}^{L_k}|\sigma_{l,k}||\sigma_{p,k}|(N_t-1)\eta_k ),
	\end{align}
	where $\eta_k \triangleq max\{ |\mathbf{R}_{k}(i,j)|\}$ and $\mathbf{R}_{k}(i,j)$ denotes the $(i,j)$-th entry of $\mathbf{R}_{k}(i,j)$.
	
	For any given local point $\tilde{\mathbf{t}}^r$ as well as lower bounds in ($\ref{C3_bound}$) and upper bounds in ($\ref{C4_bound}$), problem ($\ref{Problem4}$) is approximated as the following convex problem:
	\begin{align}
		(\text{P5})\underset{\tilde{\mathbf{t}}}{\text{max}}&~g(\tilde{\mathbf{t}}^r)+\nabla g(\tilde{\mathbf{t}})^T(\tilde{\mathbf{t}}-\tilde{\mathbf{t}}^r)-\frac{\delta}{2}(\tilde{\mathbf{t}}-\tilde{\mathbf{t}}^r)^T(\tilde{\mathbf{t}}-\tilde{\mathbf{t}}^r) \label{Problem5} \\
		\text{s.t.}
		&~\mathbf{t}_m\in \mathcal{C}_t,~1\leq m \leq N_t, \tag{\ref{Problem5}{a}} \label{Problem5a}\\		
		&~D^2 \leq -\left\| \mathbf{t}_m^r - \mathbf{t}_n^r \right\|_2^2 + 2(\mathbf{t}_m^r - \mathbf{t}_n^r)^T  \nonumber \\
		&~\times(\mathbf{t}_m - \mathbf{t}_n),~1\leq m \neq n \leq N_t, \tag{\ref{Problem5}{b}} \label{Problem5b}\\
		&~f_k(\tilde{\mathbf{t}}^r) + \nabla f_k(\tilde{\mathbf{t}}^r)^T (\tilde{\mathbf{t}} - \tilde{\mathbf{t}}^r) \nonumber \\
		&~+\frac{\zeta_k}{2}(\tilde{\mathbf{t}} - \tilde{\mathbf{t}}^r)^T(\tilde{\mathbf{t}} - \tilde{\mathbf{t}}^r) + \Gamma_k \sigma^2 \leq 0, \forall k.\tag{\ref{Problem5}{c}} \label{Problem5c}
	\end{align}
	Since the left-hand side of the constraints ($\ref{Problem5c}$) are quadratic convex with respect to $\tilde{\mathbf{t}}$, they are convex now. Furthermore, the objective function is a quadratic convex function, ($\ref{Problem5a}$), (\ref{Problem5b}), and (\ref{Problem5c}) are all linear constraints. Therefore, problem ($\ref{Problem5}$) is a convex optimization problem that can be efficiently solved with existing optimization tools.
	
	\subsection{Overall Algorithm and Convergence}
	Having obtained the solution for problem (P3) and (P5) as mentioned earlier, we can proceed with the completion of our proposed alternating optimization algorithm for solving (P1). The overall algorithm is summarized in Algorithm 1. Initially, we obtain the optimal $\mathbf{W}$ by solving (P3), while keeping $\tilde{\mathbf{t}}$ fixed. Then for given $\mathbf{W}$, we optimize FA positions by solving (P5). The algorithm iteratively solves the two sub-problems until the increase of the radar SNR in equation ($\ref{sensing_SNR}$) falls below a predefined convergence threshold $\xi$. 
	\begin{algorithm}[t]
		\caption{Alternating Optimization for Solving Problem (P1)}
		\begin{algorithmic}[1]
			\renewcommand{\algorithmicrequire}{\textbf{Input:}}
			\renewcommand{\algorithmicensure}{\textbf{Output:}}
			\REQUIRE $N_t$, $N_r$, $K$, $\mathcal{C}_t$, $\{L_k\}_{k=1}^K$, $\{ \Sigma_k \}_{k=1}^K$, $\sigma^2$, $\{\theta_{k,l}\}$, $\{\phi_{k,l}\}$, $\theta$, $\phi$, $\{\Gamma_k\}$, $P_{max}$, $D$, $\xi$.
			\STATE Initialize the FA positions $\tilde{\mathbf{t}}$. Let $r=0$. 
			\WHILE{$\text{Increase of the sensing SNR in}~(\ref{sensing_SNR})~\text{is above}~\xi$}
			\STATE Given $\tilde{\mathbf{t}}^r$, solve problem ($\ref{Problem3}$) and obtain the optimal solution by eigenvalue decomposition in (\ref{decomp}), denote the optimal solution as $\mathbf{W}^{r+1}$.
			\STATE Given $\mathbf{W}^{r+1}$, solve problem ($\ref{Problem5}$) and denote the optimal solution as $\tilde{\mathbf{t}}^{r+1}$.
			\STATE Update $r = r + 1$
			\ENDWHILE
			\ENSURE $\tilde{\mathbf{t}}_{opt}$, $\mathbf{W}_{opt}$.
		\end{algorithmic}
	\end{algorithm}
	
	The convergence analysis of Algorithm 1 is as follows. First in step 3 of Algorithm 1, given $\tilde{\mathbf{t}}^r$ the optimal solution of ($\ref{Problem2}$) can be obtained, thus we have $\gamma_r(\mathbf{W}^r,\tilde{\mathbf{t}}^r)\leq \gamma_r(\mathbf{W}^{r+1},\tilde{\mathbf{t}}^r)$. Second in step 4 of Algorithm 1, define the objective value of problem ($\ref{Problem5}$) as $g_{lb}^r(\mathbf{W},\tilde{\mathbf{t}})$, we have 
	\begin{align}
		\gamma_r(\mathbf{W}^{r+1},\tilde{\mathbf{t}}^r)&~ \overset{a}{=} g_{lb}^r(\mathbf{W}^{r+1},\tilde{\mathbf{t}}^r) \nonumber \\
		&~\overset{b}{\leq}g_{lb}^r(\mathbf{W}^{r+1},\tilde{\mathbf{t}}^{r+1}) \nonumber \\
		&~\overset{c}{\leq}\gamma_r(\mathbf{W}^{r+1},\tilde{\mathbf{t}}^{r+1}),
	\end{align}
	where ($a$) is due to the second-order Taylor expansions in ($\ref{obj_ste}$) is equal at $\tilde{\mathbf{t}}^r$; ($b$) holds since we maximize $g_{lb}^r(\mathbf{W}^{r+1},\tilde{\mathbf{t}})$ at the $r$-th iteration, equality holds when choosing $\tilde{\mathbf{t}}^{r+1}=\tilde{\mathbf{t}}^r$; ($c$) is due to the objective value of problem ($\ref{Problem5}$) serves as a lower bound of its original problem ($\ref{Problem2}$) at $\tilde{\mathbf{t}}^{r+1}$. Thus we obtain $\gamma_r(\mathbf{W}^r,\tilde{\mathbf{t}}^r)\leq \gamma_r(\mathbf{W}^{r+1},\tilde{\mathbf{t}}^{r+1})$, which indicates that the objective function of problem ($\ref{Problem}$) is non-decreasing after each iteration of Algorithm 1 and will converge.
	\section{Simulation Results}
	
	In this section, we numerically evaluate the performances of the proposed design under perfect CSI and imperfect CSI scenarios, namely Prop.P-FA and Prop.IP-FA. We set $N_t=4$, $P=Q=2$, $K=4$, $\vartheta = 45^{\circ}$, $\varphi=-30^{\circ}$, $\xi = 10^{-3}$ if not specified otherwise. The moving region for FAs is set as a square area of size $A\times A$. We set minimum SINR requirements of each communication user $\gamma_1=\cdots=\gamma_K=\gamma=10dB$. The users are randomly distributed around the BS and the distances follow uniform distributions, i.e., $d_k\sim \mathcal{U}[20,100]$, $k\in \mathcal{K}$. We set the numbers of receive paths for each user are the same, i.e., $L_k=12, k\in \mathcal{K}$. All path responses are i.i.d circularly symmetric complex Gaussian (CSCG) random variables, i.e., $\sigma_{l_k,k}\sim \mathcal{CN}(0,\rho d_k^{\alpha}/L)$, where $\rho d_k^{\alpha}$ is the expected channel power gain of user $k$, $\rho=-40$dB represents the path loss at reference distance of 1 m and $\alpha=2.8$ denotes the path loss exponent. The elevation and azimuth AoAs/AoDs are assumed to be i.i.d variables following the uniform distribution over $[\pi/2,\pi/2]$. Each point in the simulation figures is the average over 1000 user distributions and channel realizations.	 
	
	We compare the performance of the proposed with three benchmark schemes. $\textbf{(1)FPA}$: The transmitter of BS is equipped with an FPA-based uniform planner with $N_t$ antennas, spaced by $\lambda/2$. $\textbf{(2)Random position (RP)}$: Generate $\tilde{\mathbf{t}}$ satisfying ($\ref{Problema}$), ($\ref{Problemb}$) and ($\ref{Problemc}$). Obtain the radar SNR in (\ref{sensing_SNR}) by optimizing $\mathbf{W}$ with given $\tilde{\mathbf{t}}$. $\textbf{(3)Alternating position selection (APS)}$: The transmit moving region is quantized into discrete locations with equal distance $\lambda/2$. 
	
	\begin{figure}[htbp]
		\centering{\includegraphics[width=0.9\linewidth]{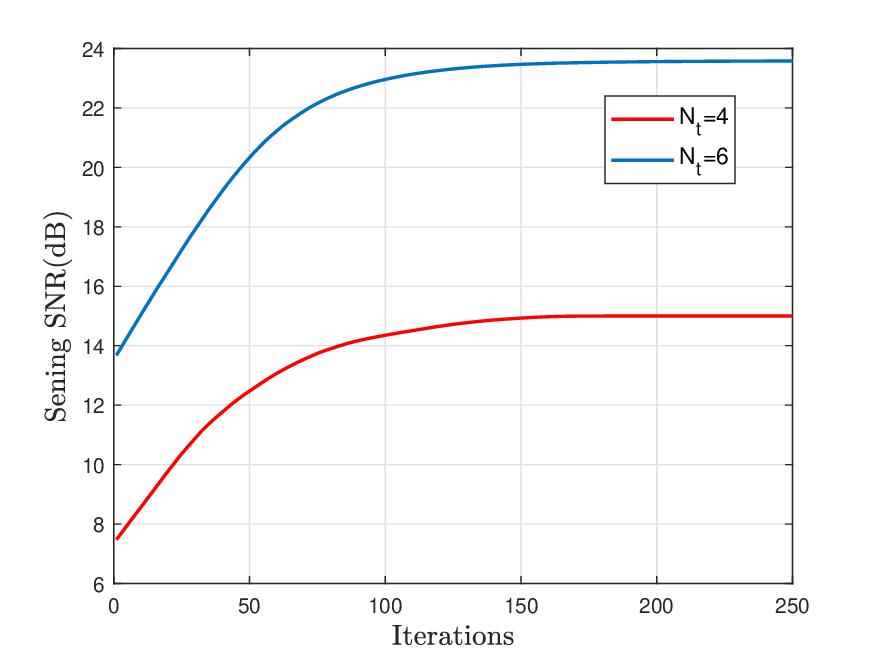}}
		\caption{ Convergence behavior of Algorithm 1.}
		\label{fig3}
	\end{figure}
	
	Fig.$\ref{fig3}$ shows the convergence behavior of Algorithm 1 with different numbers of FAs under the setup $A=2\lambda$. We can observe that for different values of $N_t$, the radar SNR increases and converges to the maximum value within about 150 iterations, which validates the convergence analysis in Section III-C. Specifically, in the case of $N_t=6$, the converged radar SNR increases by 48.2$\%$ compared to the point at 100-$th$ iteration.
	
	\begin{figure}[htbp]
		\centering{\includegraphics[width=0.9\linewidth]{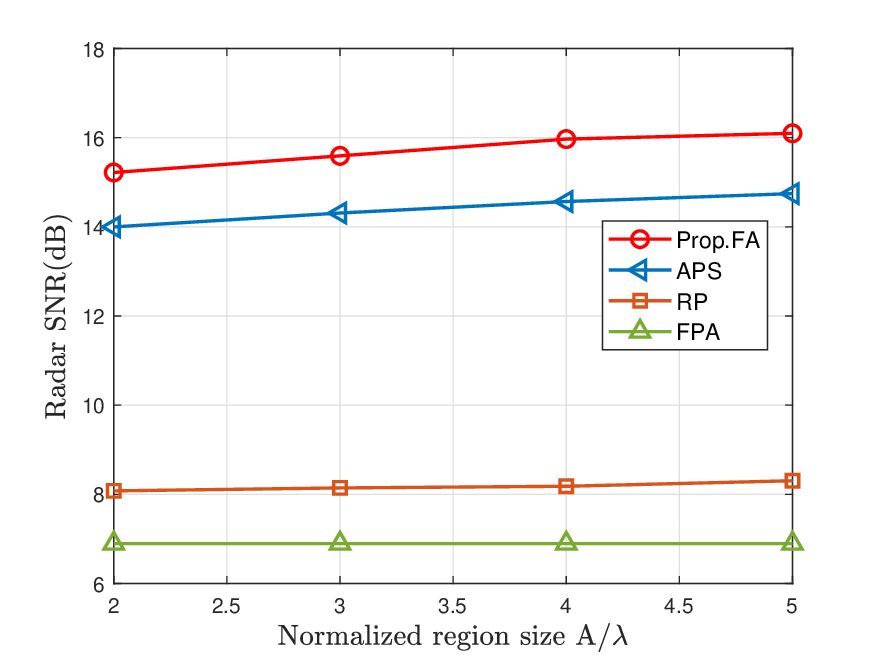}}
		\caption{ Radar SNR versus the normalized region size.}
		\label{fig1}
	\end{figure}
	Fig.$\ref{fig1}$ illustrates the radar SNR versus the normalized region size $A/\lambda$ for the Prop.FA and baselines under the setup $\gamma=10dB$. It can be observed that the radar SNRs of all designs except FPA increase with the normalized region size due to an increase in the size of the moving region, allowing further exploration of DoF in the spatial domain. Meanwhile, the proposed FA solution outperforms three baseline schemes in terms of radar SNR. Finally, when the normalized region size is greater than 5, the proposed solution converges, which indicates that the maximum radar SNR of the FA-enabled ISAC system can be achieved in a limited transmit region.
	
	\begin{figure}[htbp]
		\centering{\includegraphics[width=0.9\linewidth]{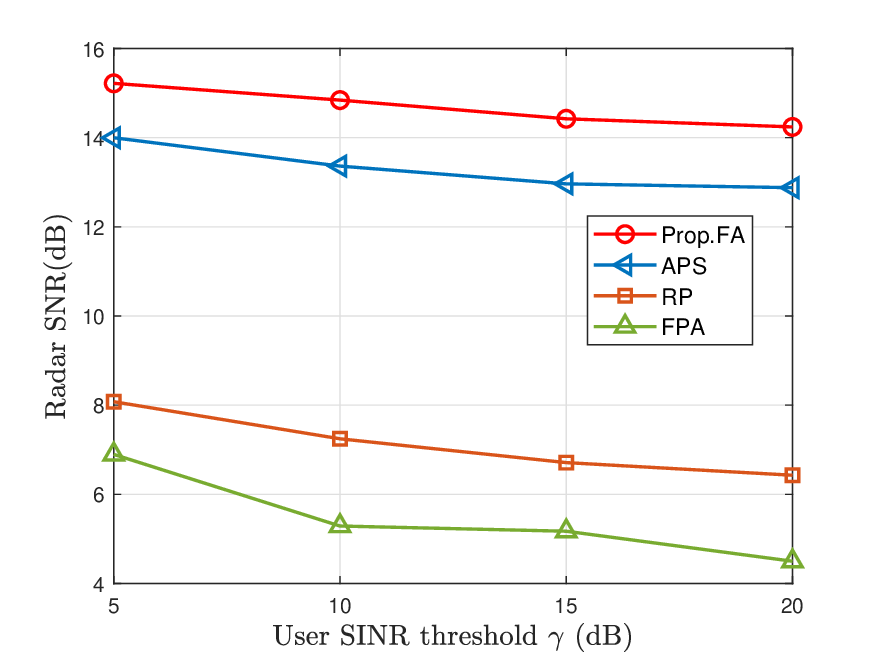}}
		\caption{ Radar SNR versus the normalized transmit region size.}
		\label{fig2}
	\end{figure}
	
	Fig.$\ref{fig2}$ shows the radar SNR versus the required threshold SINR at the user under the setup $A=2\lambda$. The radar SNR decreases with the increase of the required SINR of users for all schemes. This is attributed to the fact that the larger the user threshold SINR, the more energy the BS provides to the user. Consequently, the energy used for radar detection will decrease, which causes a decrease in radar performance. Moreover, when the threshold SINR is the same, our proposed method has a larger radar SNR compared with the three baseline schemes.
	\section{Conclusion}
	In thus paper, we investigated joint antenna position and transmit beamforming design for FA-enhanced ISAC system. First, we formulated the problem of maximizing the radar detection probability by jointly optimizing the FAs' positions and transmit beamforming matrix. Then, to tackle this non-convex problem, we proposed an AO algorithm based on SCA to obtain a sub-optimal solution. Numerical results verified that compared to FPA-based ISAC system, the proposed algorithm provided better sensing performance as well as guaranteed communication quality of service .
	\appendix
	Recall that $\tilde{\mathbf{t}}=[\mathbf{t}_1^T,...,\mathbf{t}_{N_t}^T]^T$ and $\mathbf{t}_m=[x_m,y_m]^T$, the gradient vector and Hessian matrix of $g(\tilde{\mathbf{t}})$ over $\tilde{\mathbf{t}}$ can be written as:
	\begin{equation}
		\nabla g(\tilde{\mathbf{t}}) = \left[ \frac{\partial g(\tilde{\mathbf{t}})}{\partial x_1}, \frac{\partial g(\tilde{\mathbf{t}})}{\partial y_1},..., \frac{\partial g(\tilde{\mathbf{t}})}{\partial x_{N_t}}, \frac{\partial g(\tilde{\mathbf{t}})}{\partial y_{N_t}} \right]^T,
	\end{equation}
	\begin{equation}
		\nabla^2 g(\tilde{\mathbf{t}}) =	\begin{bmatrix}
			\begin{matrix}
				\frac{\partial^2 g(\tilde{\mathbf{t}})}{\partial x_1^2} & \frac{\partial^2 g(\tilde{\mathbf{t}})}{\partial x_1 \partial y_1} \\
				\frac{\partial^2 g(\tilde{\mathbf{t}})}{\partial y_1 \partial x_1} & \frac{\partial^2 g(\tilde{\mathbf{t}})}{\partial y_1^2}
			\end{matrix}
			& \cdots & 
			\begin{matrix}
				\frac{\partial^2 g(\tilde{\mathbf{t}})}{\partial x_1 \partial y_{N_t}} \\ \frac{\partial^2 g(\tilde{\mathbf{t}})}{\partial y_1 \partial y_{N_t}}
			\end{matrix} \\
			\vdots & \ddots & \vdots \\
			\begin{matrix}
				\frac{\partial^2 g(\tilde{\mathbf{t}})}{\partial y_{N_t} \partial x_1} & \frac{\partial^2 g(\tilde{\mathbf{t}})}{\partial y_{N_t} \partial y_1}
			\end{matrix}
			& \cdots &
			\frac{\partial^2 g(\tilde{\mathbf{t}})}{\partial y_{N_t}^2}
		\end{bmatrix}.
	\end{equation}
	For simplicity, we define $\kappa_{m,n}=\rho(\mathbf{t}_m-\mathbf{t}_n)$. Combing the above, we have
	
	\begin{equation}
		\frac{\partial g(\tilde{\mathbf{t}})}{\partial x_m} = -\frac{4\pi}{\lambda}cos\theta sin\phi \sum_{n\neq m}^{N_t}|q_{m,n}|sin(\frac{2\pi}{\lambda}\kappa_{m,n}+\theta_{m,n}), \nonumber 
	\end{equation}
	\begin{equation}
		\frac{\partial g(\tilde{\mathbf{t}})}{\partial y_m} = -\frac{4\pi}{\lambda}sin\theta \sum_{n\neq m}^{N_t}|q_{m,n}|sin(\frac{2\pi}{\lambda}\kappa_{m,n}+\theta_{m,n}),
	\end{equation} 
	\begin{equation}
		\frac{\partial^2 g(\tilde{\mathbf{t}})}{\partial u_m v_m} = -\frac{8\pi^2}{\lambda}\varphi(u)\varphi(v) \sum_{n\neq m}^{N_t}|q_{m,n}|cos(\frac{2\pi}{\lambda}\kappa_{m,n}+\theta_{m,n}),\nonumber
	\end{equation}
	\begin{equation}		
		\frac{\partial^2 g(\tilde{\mathbf{t}})}{\partial u_m v_n} = \frac{8\pi^2}{\lambda}\varphi(u)\varphi(v)|q_{m,n}|cos(\frac{2\pi}{\lambda}\kappa_{m,n}+\theta_{m,n}), 
	\end{equation}
	where $1\leq m\neq n\leq N_t$, $u,v\in\{x,y\}, \varphi(x)=cos\theta sin\phi, \varphi(y)=sin\theta$.


\begin{thebibliography}{00}
		\bibitem{b1} Z. Wei, F. Liu, C. Masouros, N. Su and A. P. Petropulu, “Toward multi-functional 6G wireless networks: Integrating sensing, communication, and security,” \emph{IEEE Commun. Mag}., vol.~60, no.~4, pp. 65-71, Apr. 2022.
		\bibitem{b2} F. Liu et al. “Integrated Sensing and Communications: Toward Dual-Functional Wireless Networks for 6G and Beyond,” \emph{IEEE J. Select. Areas Commun}., vol.~40, no.~6, pp. 1728-1767, June 2022.
		\bibitem{b3} F. Liu, C. Masouros, A. P. Petropulu, H. Griffiths, and L. Hanzo, “Joint radar and communication design: Applications, state-of-the-art, and the road ahead,” \emph{IEEE Trans. Commun}., vol.~68, no.~6, pp. 3834–3862, Jun. 2020.
		\bibitem{b4} J. A. Zhang et al., “An overview of signal processing techniques for joint communication and radar sensing,” \emph{IEEE J. Sel. Topics Signal Process}., vol.~15, no.~6, pp. 1295–1315, Nov. 2021.
		\bibitem{b5} F. Liu, Y.-F. Liu, A. Li, C. Masouros, and Y. C. Eldar, “Cramér-Rao bound optimization for joint radar-communication beamforming,” \emph{IEEE Trans. Signal Process}., vol.~70, pp. 240–253, 2022.		
		\bibitem{b6} K.-K. Wong, A. Shojaeifard, K.-F. Tong, and Y. Zhang, “Fluid antenna
		systems,” \emph{IEEE Trans. Wireless Commun}., vol.~20, no.~3, pp. 1950–1962, Mar. 2021.
		\bibitem{b7} K.-K. Wong, W. K. New, X. Hao, K.-F. Tong, and C.-B. Chae, “Fluid antenna system—Part I: Preliminaries,” \emph{IEEE Commun. Lett}., vol.~27, no.~8, pp. 1919–1923, Aug. 2023.
		\bibitem{b8} K.-K. Wong, A. Shojaeifard, K.-F. Tong, and Y. Zhang, “Performance limits of fluid antenna systems,” \emph{IEEE Commun. Lett}., vol.~24, no.~11, pp. 2469–2472, Nov. 2020.
		\bibitem{b9} W. Ma, L. Zhu, and R. Zhang, “MIMO capacity characterization for movable antenna systems,” \emph{IEEE Trans. Wireless Commun}., early access, Sep. 7, 2023.
		\bibitem{b10} L. Zhu, W. Ma, B. Ning and R. Zhang, "Movable-Antenna Enhanced Multiuser Communication via Antenna Position Optimization," \emph{IEEE Trans. Wireless Commun}., early access, Dec. 12, 2023.	
		\bibitem{b11} W. K. New, K. -K. Wong, H. Xu, K. -F. Tong and C. -B. Chae, "An Information-Theoretic Characterization of MIMO-FAS: Optimization, Diversity-Multiplexing Tradeoff and q-Outage Capacity," \emph{IEEE Trans. Wireless Commun}., early access, Oct. 31, 2023.
		\bibitem{b12} K. -K. Wong, K. -F. Tong and C. -B. Chae, "Fluid Antenna System—Part III: A New Paradigm of Distributed Artificial Scattering Surfaces for Massive Connectivity," \emph{IEEE Commun. Lett}., vol.~27, no.~8, pp. 1929-1933, Aug. 2023.
		\bibitem{b13} M. Grant and S. Boyd. (2016). \emph{CVX: MATLAB Software for Disciplined Convex Programming}. [Online]. Available: http://cvxr.com/cvx.
		\bibitem{b14} Y. Sun, P. Babu and D. P. Palomar, "Majorization-Minimization Algorithms in Signal Processing, Communications, and Machine Learning," \emph{IEEE Trans, Signal Process}., vol.~65, no.~3, pp. 794-816, Feb. 2017.
		
	\end{thebibliography}
\end{document}